\documentclass[aps,a4paper,english,nofootinbib,showpacs,preprint]{revtex4}
\usepackage{color}
\usepackage{amssymb}
\usepackage{amsmath}
\usepackage{graphicx}
\usepackage[T1]{fontenc}
\usepackage{hyperref}
\usepackage[latin1]{inputenc}
\usepackage{amsfonts}

\begin{document}

\title{Wide Localized Solitons in Systems with Time and Space-Modulated
Nonlinearities}
\author{L.E. Arroyo-Meza}
\email{luisarroyo@feg.unesp.br}
\author{A. de Souza Dutra}
\email{dutra@feg.unesp.br}
\author{M. B. Hott}
\email{marcelo.hott@pq.cnpq.br}
\affiliation{UNESP Universidade Estadual Paulista - Campus de Guaratinguetá - DFQ.\\
Av. Dr. Ariberto Pereira da Cunha, 333 CEP 12516-410,
Guaratinguetá-SP, Brazil.} 

\begin{abstract}
In this work we apply point canonical transformations to solve some
classes of nonautonomous nonlinear Schr\"{o}dinger equation namely,
those which possess specific cubic and quintic - time and space
dependent - nonlinearities. In this way we generalize some
procedures recently published which resort to an ansatz to the
wavefunction and recover a time and space independent nonlinear
equation which can be solved explicitly. The method applied here
allow us to find wide localized (in space) soliton solutions to the
nonautonomous nonlinear Schr\"{o}dinger equation, which were not
presented before. We also generalize the external potential which
traps the system and the nonlinearities terms.
\end{abstract}

\maketitle

\section{Introduction}

Investigations on the nonlinear Schr\"{o}dinger equation (NLSE)
\cite{sulem} has increased very much in the last decades. The
importance of such investigations are not only due the possible real
world applications of NLSE or the Gross-Pitaevskii equation (GPE)
\cite{pita1}, but also due to the possibility, on the theoretical
side, in increasing the class of nonlinear integrable models
\cite{das}.

The applications of NLSE and/or GPE with spatially dependent cubic
and quintic (CQ) nonlinearities can be appreciated, for example, in
pulse
propagation in optical fibers \cite{appl1} in photonic crystals \cite{appl2}%
, and in the study of Bose-Einstein (BEC) condensates
\cite{applbec}, whose nonlinearities are driven by means of optical
interactions as well.

Especifically, dark and bright solitons have been observed in
various nonlinear physical phenomena. Investigation of bright and
dark solitons is useful for understanding the properties of BEC.
Bright solitons are characterized by a localized maximum in the
density profile without any phase jump across it. In the relevant
experiments, this type of soliton is formed upon utilizing a
Feshbach resonance to change the sign of the scattering length from
positive to negative. On the other hand, dark solitons may also be
considered as moving domain walls\ which separate regions of a
condensate with different values of the order parameter. In fact,
dark solitons are density dips characterized by a phase jump of the
wave function at the position of the dip and can be generated by
means of phase-engineering techniques \cite{Burger}. Thus, there are
many possibilities of managing these solitons, strongly justifying a
quest for novel analytical solutions.

The thecniques for managing the nonlinearities have been improved a
lot and, in some cases, the nonlinear equations governing the system
present not only
space \cite{sdependent} but time-dependent nonlinearities also \cite%
{tdependent}-\cite{malomed}. In this case one talks about the
so-called nonautonomous NLSE \cite{serkin1}, whose localized waves
solutions, in some specific cases, were found by Serkin and Hasegawa
\cite{serkin2} resorting to a similarity transformation, which maps
the nonautonomous NLSE onto a nonlinear stationary equation, whose
solutions are well known.

Similar procedure has been carried out in the recent study by Beitia, P\'{e}%
rez-Garc\'{\i}a, Vekslerchik and Konotop (BPVK) \cite{BPVK}.
Concretely,\
the authors relate the nonautonomous NLSE with cubic nolinearity (CNLSE)%
\begin{equation}
i\frac{\partial \Psi }{\partial t}=-\frac{\partial ^{2}\Psi }{\partial x^{2}}%
+v\left( x,t\right) \Psi +g\left( x,t\right) \left\vert \Psi
\right\vert ^{2}\Psi ,  \label{1}
\end{equation}%
and the stationary CNLSE with constant coefficients
\begin{equation}
\mu \Phi =-\frac{\partial ^{2}\Phi }{\partial X^{2}}+G\left\vert
\Phi \right\vert ^{2}\Phi ,  \label{2}
\end{equation}%
($\mu $\ is the eigenvalue of the nonlinear equation known as
chemical
potential) by using the ansatz%
\begin{equation}
\Psi \left( x,t\right) =\rho \left( x,t\right) \exp \left( i\varphi
\left( x,t\right) \right) \Phi \left( X\left( x,t\right) \right) .
\label{3}
\end{equation}%
for the solution of Eq. (\ref{1}). The variable $X$, which in Eq.
(\ref{2}) plays the role of space coordinate, is in fact a
function$~F\left( \xi \right) $ of the specific combination $\xi
\left( x,t\right) =\gamma (t)x+\delta (t)$ of the space coordinate
and the time. In such an approach, which provides analytical
solutions for the wavefunction $\Psi \left(
x,t\right) $, one finds out that specific forms of the trapping potential $%
v\left( x,t\right) $ and of the nonlinearity function $g\left(
x,t\right) $ can be explored and that there is an intrinsic
dependence of $v\left( x,t\right) $, $g\left( x,t\right) $, $\rho
\left( x,t\right) $ and $\varphi
\left( x,t\right) $ on $\gamma (t)$, $\delta (t)$ and\ $F\left( \xi \right) $%
, that is one may choose conveniently $\gamma (t)$, $\delta (t)$\ and $%
F\left( \xi \right) $ to obtain $v\left( x,t\right) $, $g\left(
x,t\right) $ and, consequently nonsingular $\rho \left( x,t\right) $
and single-valued phase $\varphi \left( x,t\right) $.

More recently, in the work by Avelar, Bazeia and Cardoso (ABC) \cite{ABC}%
\thinspace\ the authors followed the BPVK approach but extend the
problem, focusing their attention on the nonautonomous cubic and
quintic nonlinear Schr\"{o}dinger equation (CQNLSE). They have also
obtained analytic localized solutions of the bright or dark type
(breathing, resonant, quasiperiodic and moving breathing solutions),
depending on whether the eigenvalue $\mu ~$vanishes or not.

The nonautonomous CQNLSE is obtained by adding a term $h\left(
x,t\right) \left\vert \Psi \right\vert ^{4}\Psi $ to the right side
of (\ref{2}) and
the stationary CQNLSE with constant coefficients is given by%
\begin{equation}
\mu \Phi =-\frac{\partial ^{2}\Phi }{\partial X^{2}}+G_{3}\left\vert
\Phi \right\vert ^{2}\Phi +G_{5}\left\vert \Phi \right\vert ^{4}\Phi
.  \label{4}
\end{equation}

We have noticed that when the similarity transformation is applied,
the trapping potential $v\left( x,t\right) $ always has a quadratic
term, namely $\omega ^{2}(t)x^{2}$, which constitutes a harmonic
oscillator with time-dependent frequency. The frequency is related
to $\gamma (t)$ and its time derivatives and, when $d^{2}\delta
/dt^{2}\neq 0$, one finds a driven time-dependent harmonic
oscillator, whose force is also time-dependent. Such
kind of problem is consacred in the literature of the time-dependent Schr%
\"{o}dinger equation concerning the analysis of dissipative effects
in quantum fluctuations. Moreover, such systems have found real
world
applications in quantum optics \cite{colegrave} and plasma physics \cite{ben}%
. Among the approaches applied to solve the Schr\"{o}dinger equation
for the time-dependent (driven) oscillator one resorts to point
canonical transformations on the coordinates and a re-scaling of the
time in such a way the problem can be transformed into a
Schr\"{o}dinger equation for the harmonic oscillator with constant
frequency \cite{farinadutra}. As a matter of fact, in some specific
cases, the problem can be even reduced further into a
Schr\"{o}dinger equation for a free particle \cite{jackiw}.

One of the main goals of the present work is to apply successfully
the same
canonical point \ transformations for solving some time-dependent Schr\"{o}%
dinger equations to the problem of the nonautonomous CQNLSE
mentioned above. We show that the point canonical transformation
followed by an appropriate redefinition of the wavefunction and
additional transformations of variables also leads to the stationary
CQNLSE with constant coefficients (\ref{4}), without resorting to
any ansatz on the form of the wave function $\Psi \left( x,t\right)
$ in Eq. (\ref{3}).

The approach adopted here is straightforward and as a result of it
one can see clearly how is the dependence of the trapping potential
$v\left( x,t\right) $ and of the inhomogeneous coefficients $g\left(
x,t\right) $ and
$h\left( x,t\right) $ with the functions $\gamma (t)$, $\delta (t)$\ and $%
F\left( \xi \right) $. With this approach, the constraints over the
functions $\gamma (t)$, $\delta (t)$\ and $F\left( \xi \right) ~$in
order to render the wavefunction $\Psi \left( x,t\right) $ well
defined are also evident. By following the mapping approach
presented in \cite{dhb} we present the soliton solutions for Eq.
(\ref{4}) in terms of the Weierstrass
elliptic function, such that by considering specific sets of the parameters $%
\mu $, $G_{3},~G_{5}$ and an arbitrary constant of integration we
not only recover the results found in \cite{BPVK}-\cite{ABC}, but
also present some new soliton solutions by using the set of
functions $\gamma (t)$, $\delta (t) $\ and $F\left( \xi \right) $
they have worked with.

We also deal with extensions of the models by including other
trapping potentials which are a mixing of circular functions with
the time-dependent harmonic oscillator and also comment on a
generalization of the nonautonomous time-dependent nonlinear
Schr\"{o}dinger equation by considering a non-polynomial
nonlinearity.

In the next section we present the approach to map the nonautonomous
CQNLSE onto a stationary CQNLSE. The third section is devoted to
present the solutions of (\ref{4}) in terms of the Weierstrass
elliptic function where some examples of wide bright and dark
solitons are shown. In the fourth section we consider other kinds of
trapping potential besides the persistent time-dependent harmonic
oscillator and comment on a non-polynomial nonlinearity. The fifth
section is left for the conclusions.

\section{The approach}

In this section we present the approach by focusing on the
nonautonomous nonlinear Schr\"{o}dinger equation with terms of cubic
and quintic order in the wavefunction, namely
\begin{equation}
i\frac{\partial \Psi }{\partial t}=-\frac{\partial ^{2}\Psi }{\partial x^{2}}%
+v\left( x,t\right) \,\Psi +g_{3}\left( x,t\right) \,|\Psi
|^{2}\,\Psi +g_{5}\left( x,t\right) \,|\Psi |^{4}\,\Psi ,
\label{eq1}
\end{equation}

\noindent whose function coefficients are written as
\begin{equation}
v\left( x,t\right) =\omega \left( t\right) x^{2}+f_{1}\left(
t\right) x+f_{2}\left( t\right) +\gamma ^{2}\left( t\right)
\,V\left[ \gamma \left( t\right) \,x+\delta \left( t\right) \right]
\,,  \label{eq2}
\end{equation}%
\begin{equation}
g_{3}\left( x,t\right) =G_{3}\,\gamma \left( t\right) \,f\left[
\gamma \left( t\right) \,x+\delta \left( t\right) \right] \,,
\label{eq3}
\end{equation}%
\begin{equation}
g_{5}\left( x,t\right) =G_{5}\,h\left[ \gamma \left( t\right)
\,x+\delta \left( t\right) \right] \,.  \label{eq4}
\end{equation}%
The reason for choosing the inhomogeneous coefficients in this way
is going to be clarified below.

Now, we implement the following coordinate transformation and time
rescaling \cite{farinadutra}
\begin{equation}
x=\frac{\xi }{\overline{\gamma }\left( \tau \right)
}-\frac{\overline{\delta }\left( \tau \right) }{\overline{\gamma
}\left( \tau \right) }\,, \label{eq5}
\end{equation}%
\begin{equation}
t-t_{0}=\int_{0}^{\tau }\frac{d\tau ^{\prime }}{\overline{\gamma
}^{2}\left( \tau ^{\prime }\right) }\,,  \label{eq6}
\end{equation}

\noindent with $\overline{\gamma }\left[ \tau \left( t\right)
\right]
=\gamma \left( t\right) $ and $\overline{\delta }\left[ \tau \left( t\right) %
\right] =\delta \left( t\right) \,$. Then, one can recast
(\ref{eq1}) as
\begin{eqnarray}
&&\left. i\,\overline{\gamma }\left( \overline{\gamma }_{\tau }\,\xi -%
\overline{\gamma }_{\tau }\,\overline{\delta }+\overline{\gamma }\,\overline{%
\delta }_{\tau }\right) \frac{\partial \overline{\Psi }}{\partial \xi }+i\,%
\overline{\gamma }^{2}\,\frac{\partial \overline{\Psi }}{\partial \tau }%
=\right.  \nonumber \\
&&\left. =-\overline{\gamma }^{2}\,\frac{\partial ^{2}\overline{\Psi }}{%
\partial \xi ^{2}}+\overline{\upsilon }\left( \xi ,\tau \right) \,\overline{%
\Psi }+G_{3}\,\overline{\gamma }\,\overline{f}\left( \xi \right) |\overline{%
\Psi }|^{2}\,\overline{\Psi }+G_{5}\,\overline{h}\left( \xi \right) |%
\overline{\Psi }|^{4}\,\overline{\Psi }\right. \,,  \label{7}
\end{eqnarray}

\noindent where $\overline{\gamma }_{\tau }=d\overline{\gamma }/d\tau $%
\thinspace ,\thinspace $\overline{\delta }_{\tau }=d\overline{\delta
}/d\tau $\thinspace ,\thinspace $\overline{\Psi }\left( \xi ,\tau
\right) =\Psi \left( x\left( \xi ,\tau \right) ,t\left( \tau \right)
\right) $\thinspace\ and
\begin{equation}
\overline{\upsilon }\left( \xi ,\tau \right) =\overline{\omega
}\left( \tau
\right) \frac{\left( \xi -\overline{\delta }\right) ^{2}}{\overline{\gamma }%
^{2}}+\overline{f}_{1}\left( \tau \right) \frac{\left( \xi
-\overline{\delta
}\right) }{\overline{\gamma }}+\overline{f}_{2}\left( \tau \right) +%
\overline{\gamma }^{2}\left( \tau \right) \,V\left( \xi \right) \,.
\label{eq8}
\end{equation}%
From the last two equations one can appreciate why we have chosen
the specific dependence of $V\,$, $f$ and $h$ on $\xi =\gamma \left(
t\right) \,x+\delta \left( t\right) $. Eq. (\ref{7}) looks like a
nonautonomous NLSE
on $\xi $ and $\tau $, except for the first derivative term in the variable $%
\xi $. In order to eliminate that term, we redefine $\overline{\Psi
}\left( \xi ,\tau \right) $ as
\begin{equation}
\overline{\Psi }=\sqrt{\overline{\gamma }\left( \tau \right) }\,\mathrm{e}%
^{-i\,\overline{\alpha }\left( \xi ,\tau \right) }\,\psi \left( \xi
,\tau \right) \,,  \label{eq9}
\end{equation}

\noindent where $\overline{\alpha }\left( \xi ,\tau \right) =\frac{\overline{%
\gamma }_{\tau }}{4\,\overline{\gamma }}\xi ^{2}+\left( \overline{\delta }%
_{\tau }-\frac{\overline{\gamma }_{\tau }\overline{\delta }}{\overline{%
\gamma }}\right) \frac{\xi }{2}-\overline{a}\left( \tau \right) \,$, with $%
\overline{a}\left( \tau \right) $ an arbitrary function for the
moment. By substituting (\ref{eq9}) in (\ref{7}) one gets
\begin{equation}
i\,\frac{\partial \psi }{\partial \tau }=-\,\frac{\partial ^{2}\psi }{%
\partial \xi ^{2}}+U\left( \xi ,\tau \right) \psi +G_{3}\,f\left( \xi
\right) |\psi |^{2}\,\psi +G_{5}\,h\left( \xi \right) |\psi
|^{4}\,\psi \,, \label{eq10}
\end{equation}%
where
\begin{eqnarray*}
U\left( \xi ,\tau \right) &=&\left. \left( \overline{\omega }\left(
\tau
\right) -\frac{\overline{\gamma }^{3}\,\overline{\gamma }_{\tau \tau }}{4}%
\right) \frac{\left( \xi -\overline{\delta }\right) ^{2}}{\overline{\gamma }%
^{4}}+\left( \overline{f}_{1}\left( \tau \right) -\frac{\overline{\gamma }%
^{3}\,\overline{\delta }_{\tau \tau }}{2}\right) \frac{\left( \xi -\overline{%
\delta }\right) }{\overline{\gamma }^{3}}\right. \\
&&\left. +\frac{1}{\overline{\gamma }^{2}}\left(
\overline{f}_{2}\left( \tau
\right) +\frac{\overline{\delta }_{\tau }^{2}\,\overline{\gamma }^{2}}{4}+%
\overline{\gamma }^{2}\frac{d\overline{a}}{d\tau }\right) +\,V\left(
\xi \right) .\right.
\end{eqnarray*}%
$\,$One can see why the factors involving $\gamma \left( t\right) $
are present in the expressions of $v\left( x,t\right) $ and
$g_{3}\left( x,t\right) $. Moreover, one can also appreciate the
contribution of the
redefinition (\ref{eq9}), specifically the contribution of the phase $%
\overline{\alpha }\left( \xi ,\tau \right) $ to the redefinition of
the trapping potential $U\left( \xi ,\tau \right) $.

Now, one can choose $\gamma \left( \tau \right) $ and $\delta \left(
\tau \right) $ such that
\begin{equation}
\overline{\omega }\left( \tau \right) =\frac{\overline{\gamma }^{3}\,%
\overline{\gamma }_{\tau \tau }}{4}\,,~\overline{f}_{1}\left( \tau \right) =%
\frac{\overline{\gamma }^{3}\,\overline{\delta }_{\tau \tau }}{2}\,,~%
\overline{f}_{2}\left( \tau \right) =-\overline{\gamma }^{2}\left( \frac{%
\overline{\delta }_{\tau }^{2}}{4}+\frac{d\overline{a}}{d\tau
}\right) \, \label{eq11}
\end{equation}%
and, by coming back to the original variables ($x,t$), the functions
$\omega $, $f_{1}$ and $f_{2}$ appear as
\begin{equation}
\omega \left( t\right) =\frac{\gamma _{tt}\gamma -2\gamma
_{t}^{2}}{4\gamma ^{2}}\,,\qquad f_{1}\left( t\right) =\frac{\delta
_{tt}\gamma -2\gamma _{t}\delta _{t}}{2\gamma ^{2}}\,,\qquad
f_{2}\left( t\right) =-\frac{\delta _{t}^{2}}{4\gamma
^{2}}-\frac{da}{dt}\,,  \label{eq12}
\end{equation}%
revealing the intrinsic connection between the frequency $\omega
\left( t\right) $, the force $f_{1}\left( t\right) $ and the
functions $\gamma \left( t\right) $ and $\delta \left( t\right) $.
Note that $a(t)$ is an arbitrary function that could be chosen as
$da/dt=(\delta _{t}/2\gamma )^{2}$ if $f_{2}\left( t\right) $ were
not present in (\ref{eq2}). Thus, one removes the explicit
time-dependency of (\ref{eq10}), that is

\begin{equation}
i\,\frac{\partial \psi }{\partial \tau }=-\frac{\partial ^{2}\psi
}{\partial \xi ^{2}}+V\left( \xi \right) \psi +G_{3}\,f\left( \xi
\right) |\psi |^{2}\,\psi +G_{5}\,h\left( \xi \right) |\psi
|^{4}\,\psi \,.  \label{eq13}
\end{equation}

\noindent and the wavefunction (\ref{eq9}) is written as%
\begin{equation}
\Psi \left( x,t\right) =\sqrt{\gamma \left( t\right) }\,\mathrm{e}%
^{-i\,\alpha \left( x,t\right) }\,\psi \left( \xi \left( x,t\right)
,\tau \left( t\right) \right) \,,  \label{eq9a}
\end{equation}

\noindent where $\alpha \left( x,t\right) =\frac{\gamma _{t}}{4\,\gamma }%
\,x^{2}+\frac{\delta _{t}}{2\,\gamma }\,x-a\left( t\right) $. We
recall that the phase $\alpha \left( x,t\right) $ was introduced
through the redefinition (\ref{eq9}) and it contributes to the
redefinition of the trapping potential, which could be eliminated
thanks to the presence of the
time-dependent driven harmonic oscillator terms (the three first terms $%
v\left( x,t\right) )$.

For stationary solutions in the variable $\tau $, that is $\psi (\xi
,\tau )=\,\phi \left( \xi \right) \exp (-\,iE\,\tau )$, we have
\begin{equation}
\frac{d^{2}\phi }{d\xi ^{2}}=\left( V\left( \xi \right) -E\right)
\phi +G_{3}\,f\left( \xi \right) |\phi |^{2}\,\phi +G_{5}\,h\left(
\xi \right) |\phi |^{4}\,\phi \,.  \label{eq14}
\end{equation}

Since we still have a nonlinear equation with inhomogeneous
nonlinearities, we are going to make further transformations in
order to reach a nonlinear second-order differential equation with
constant couplings. For that we
redefine $\xi $ as a function of another variable $\zeta $, that is $\xi =%
\overline{F}\left( \zeta \right) $, \noindent which lead us to the
differential equation
\begin{eqnarray}
&&\left. \frac{d^{2}\phi }{d\zeta ^{2}}-\frac{\overline{F}_{\zeta \zeta }}{%
\overline{F}_{\zeta }}\frac{d\phi }{d\zeta }=\overline{F}_{\zeta
}^{2}\left( V\left[ \xi \left( \zeta \right) \right] -E\right) \phi
\right.   \nonumber
\\
&&\left. +G_{3}\,\overline{F}_{\zeta }^{2}\,f\left[ \xi \left( \zeta \right) %
\right] |\phi |^{2}\,\phi +G_{5}\,\overline{F}_{\zeta }^{2}\,h\left[
\xi \left( \zeta \right) \right] |\phi |^{4}\,\phi \right. ,
\end{eqnarray}%
where $\overline{F}_{\zeta }=d\overline{F}/d\zeta $ and $\phi =\phi
\left( \xi \left( \zeta \right) \right) $. By redefining the field
as
\begin{equation}
\phi \left( \zeta \right) =\sqrt{\overline{F}_{\zeta }}\,\Phi \left(
\zeta \right) \,,  \label{eq17}
\end{equation}

\noindent we reach Eq. (\ref{eq4}) in terms of $\zeta $
\begin{equation}
\frac{d^{2}\Phi }{d\zeta ^{2}}=-\mu \,\Phi +G_{3}|\Phi |^{2}\,\Phi
+G_{5}\,|\Phi |^{4}\,\Phi \,,  \label{eq18}
\end{equation}

\noindent where
\begin{equation}
\mu =-\overline{F}_{\zeta }^{2}\left( V\left[ \xi \left( \zeta \right) %
\right] -E\right) +\frac{\overline{F}_{\zeta \zeta \zeta }}{2\overline{F}%
_{\zeta }}-\frac{3\overline{F}_{\zeta \zeta }^{2}}{4\overline{F}_{\zeta }^{2}%
}\,,~~f\left[ \xi \left( \zeta \right) \right]
=\frac{1}{\overline{F}_{\zeta
}^{3}}\,,~h\left[ \xi \left( \zeta \right) \right] =\frac{1}{\overline{F}%
_{\zeta }^{4}}\,.
\end{equation}%
\newline

\noindent In terms of the variables $\xi $ and $\tau $ we have%
\begin{equation}
V\left( \xi \right) =\left( \frac{F^{\prime \prime }}{2F^{\prime
}}\right) ^{2}-\left( \frac{F^{\prime \prime }}{2F^{\prime }}\right)
^{\prime }-\mu F^{\prime }{}^{2}+E\,,~f\left( \xi \right) =F^{\prime
^{3}}{},~h\left( \xi \right) =F^{\prime ^{4}}{}\,,  \label{eq18b}
\end{equation}

\noindent where $F^{\prime }=dF/d\xi \,$ and the field (\ref{eq17}) becomes%
\begin{equation}
\phi \left( \xi \right) =\frac{1}{\sqrt{F^{\prime }\left( \xi
\right) }}\Phi \left( \zeta \left( \xi \right) \right) .
\label{eq18a}
\end{equation}

Finally, by returning to the original space-time coordinates
($x,t$), the wavefunction can be obtained from (\ref{eq9a}) and
(\ref{eq18a}), that is

\begin{equation}
\Psi \left( x,t\right) =\frac{\sqrt{\gamma \left( t\right) }}{\sqrt{%
F^{\prime }\left[ \xi \left( x,t\right) \right] }}\,\exp \left[
-i\,\eta
\left( x,t\right) \right] \Phi \left[ F\left( \xi \left( x,t\right) \right) %
\right] ,  \label{foxt}
\end{equation}%
with $\eta \left( x,t\right) =\frac{\gamma _{t}}{4\,\gamma }x^{2}+\frac{%
\delta _{t}}{2\,\gamma }x-a\left( t\right) +E\int_{0}^{t}dt^{\prime
}\gamma ^{2}\left( t^{\prime }\right) $, where $a(t)$ is an
arbitrary function.

Thus, we have shown, by means of transformation of variables, how
the nonautonomous CQNLSE, Eq. (\ref{eq1}), can be mapped onto a
nonlinear one-dimensional second-order differential equation with
cubic and quintic nonlinearities Eq. (\ref{eq18}). More than that,
we have shown explicitly how the part of the trapping potential
$V(\xi )$ and the nonlinearities functions $f(\xi )$ and $h(\xi
)~$are related to the transformation function $F(\xi
)=\overline{F}^{-1}(\xi )$ (see eqs. (\ref{eq18b})).

\section{The mapping onto the Weierstrass $\wp $-function}

Here we consider the cases in which $\Phi $ is a real function on $\zeta $ (%
\ref{eq18}). One can check that the solutions for eq. (\ref{eq18})
are also solutions of the equation
\begin{equation}
\left( \frac{d\Phi }{d\zeta }\right) ^{2}=\epsilon -\mu \,\Phi ^{2}+\frac{%
G_{3}}{2}\Phi ^{4}+\frac{G_{5}}{3}\Phi ^{6}\,.  \label{eq19}
\end{equation}%
where $\epsilon $ is a real arbitrary constant. Moreover, we shown
how to map such kind of equations \cite{dhb} onto the nonlinear
Weierstrass differential equation
\begin{equation}
\left( \frac{d\wp }{d\zeta }\right) ^{2}=4\,\wp ^{3}-g_{2}\,\wp
-g_{3}\,, \label{w}
\end{equation}%
where $g_{2}$ and $g_{3}$ are the Weierstrass invariants and the
discriminant is $\Delta =g_{2}^{3}-27g_{3}^{2}$\thinspace . The
values of the invariants and of the discriminant determine how the
Weierstrass function are written in terms of the double-periodic
Jacobi elliptic
functions for $\Delta \neq 0$ \cite{abramowitz}, and the solutions for (\ref%
{eq19}) can be included in one of the categories listed below

\textit{1)} For $\epsilon \neq 0$ one has the following solution for
$\Phi \left( \zeta \right) $
\begin{equation}
\Phi \left( \zeta \right) =\sqrt{\frac{\epsilon }{\wp \left( \zeta
,g_{2},g_{3}\right) +\mu /3}}\,,  \label{eq20}
\end{equation}%
where $g_{2}=\frac{4}{3}\mu ^{2}-2\,G_{3}\,\epsilon \,$ and $\,g_{3}=\frac{8%
}{27}\mu ^{3}-\frac{2}{3}G_{3}\,\mu \,\epsilon
-\frac{4}{3}G_{5}\,\epsilon
^{2}\,$ which leads to the following discriminant $\Delta =-\frac{%
4\,\epsilon ^{2}}{3}\left( 6\epsilon \,G_{3}^{3}+36\epsilon
^{2}\,G_{5}^{2}+36G_{3}G_{5}\epsilon \,\mu -3\mu ^{2}G_{3}^{2}-16\mu
^{3}G_{5}\right) \,.$

\noindent \qquad \textit{2) }For $\epsilon =0$ and $\mu \neq 0$ one
sees
that $\,g_{2}=\frac{4}{3}\mu ^{2}\,$, $g_{3}=\frac{8}{27}\mu ^{3}\,$ and $%
\,\Delta =0$, then the solution is
\begin{equation}
\Phi \left( \zeta \right) =\left( \frac{G_{3}-2\beta }{4\mu }+\frac{\beta }{%
\wp \left( \zeta ,g_{2},g_{3}\right) +\mu /3}\right) ^{-1/2}\,,
\label{eq21}
\end{equation}%
where $\beta =\pm \sqrt{\frac{G_{3}^{2}}{4}+\frac{4}{3}\mu
\,G_{5}}\,$.

\noindent \qquad \textit{3) }For $\epsilon =0$ and $\mu =0$ one gets $%
g_{2}=g_{3}=\Delta =0$ (in this case $\wp \left( \zeta \right) =\zeta ^{-2}$%
) and
\begin{equation}
\Phi \left( \zeta \right) =\left( -\frac{2\,G_{5}}{3\,G_{3}}+\frac{%
G_{3}\zeta ^{2}}{2\,}\right) ^{-1/2}\,,  \label{eq22}
\end{equation}

All the examples with real world applications considered in
\cite{BPVK} and \cite{ABC} may be reproduced and fit in one of the
these categories. We present next three additional examples which
were not considered by them.

\subsection{Examples}

Because the dark and bright solitons are of primary importance for
developing concrete applications of BEC, many technique are
developed for manipulating and controlling the soliton's parameters
and induce changes in their shapes which would be useful for
applications. One possibility is to vary the atomic scattering
length by means of external magnetic fields, i.e. by using Feshbach
resonances. People has been asked to themselves on how to modulate
the width and amplitude of a soliton in a controllable manner. It
has been demonstrated that the variation of the scattering length
provides a powerful tool for controlling the generation of bright
and dark soliton trains ; especially, this technique can be used to
modulate the bright soliton into very high local matter densities in
both harmonic trap potential \cite{salerno} and repulsive (inverted)
harmonic trap potential \cite{zhang}.

We show by means of some examples that the width and the amplitude
of the solitons can also be manipulated theoretically by focusing on
the same form of the functions $F\left( \xi \right) ,\gamma \left(
t\right) $ and $\delta \left( t\right) $ considered in \cite{ABC}
but with some different values of $\mu ,$ $\,G_{3},$ $\,G_{5}$ and
$\epsilon $. Explicitly, we take $F\left( \xi \right)
=\frac{\sqrt{3\,\pi }\,b}{2\,G_{3}^{1/3}}\mathrm{Erfi}\left(
\frac{\xi }{\sqrt{3}\,b}\right) $ (this is the imaginary error function \cite%
{abramowitz}), $\gamma \left( t\right) =\sqrt{\frac{2}{1+3\cos
^{2}\left( 2t\right) }}$ and $\delta \left( t\right) =0$. We remark
that this choice of $F\left( \xi \right) $, when substituted in
$V(\xi )$ (\ref{eq18b}), implies into an extra contribution for the
time-dependent frequency $\omega (t)$ of
the driven harmonic oscillator, or in other words, the trapping potential $%
v\left( x,t\right) $ is a sole time-dependent driven harmonic
oscillator for $\mu =0$.

\subsubsection{\textit{Example 1:}}

In real world applications the nonlinearities and the confinement
potential can be changed independently using an optical trap and
magnetic-field-induced Feshbach resonance. In order to modulate the
width of the dark soliton in a controllable manner, we gradually
increase the self-interaction and simultaneously turn off the
trapping potential.

Here, there is no way to turn off the ubiquitous trapping harmonic
potential, but we can modulate the width of a dark soliton through a
suitable choice of the chemical potential and the coefficient of the
quartic self-interactions. Here, as in real world applications, they
are related to each other.

First we notice that Eq. (\ref{eq18}) can be seen as the static
equation of motion $d^{2}\Phi /d\zeta ^{2}=dU/d\Phi $, where $U(\Phi
)=(1/2)(-\mu \Phi ^{2}\,+\frac{G_{3}}{2}\Phi
^{4}+\frac{G_{5}}{3}\,\Phi ^{6}+c)$. If the parameters of the
potential are chosen such that it presents only two minima, say at
$\pm 1$, and such that the chemical potential varies from positive
to negative values (we notice that $d^{2}U/d\Phi _{\Phi =0}^{2}=-\mu
$), the minimum energy solution $\Phi (\zeta )$, which connects the
minima of $U(\Phi )$ at $\zeta \rightarrow \pm \infty $, is a kink
which is deformable into a two-kink (double kink) as $\mu $ varies
from positive to negative values.

Such $\Phi ^{6}$ polynomial potentials have been used to study phase
transitions which comes with domain wall splitting and the
appearance of a wet phase in some ferroelectric \cite{dorfman} and
paramagnetic \cite{gordon} materials. The wetting transition can
also take place in deconfinement phase transition of SU(3)
Yang-Mills theory \cite{frei}, in supersymmetric QCD \cite{wiese}
and in thick brane-world scenario \cite{campos}, \cite{hott}.

\begin{figure}[h]
\centering
\includegraphics[width=7.0cm]{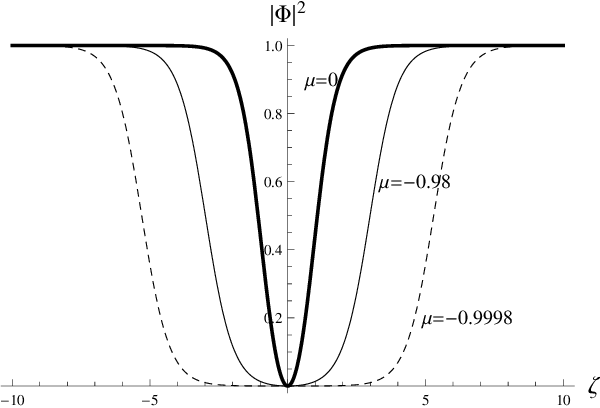}
\includegraphics[width=7.0cm]{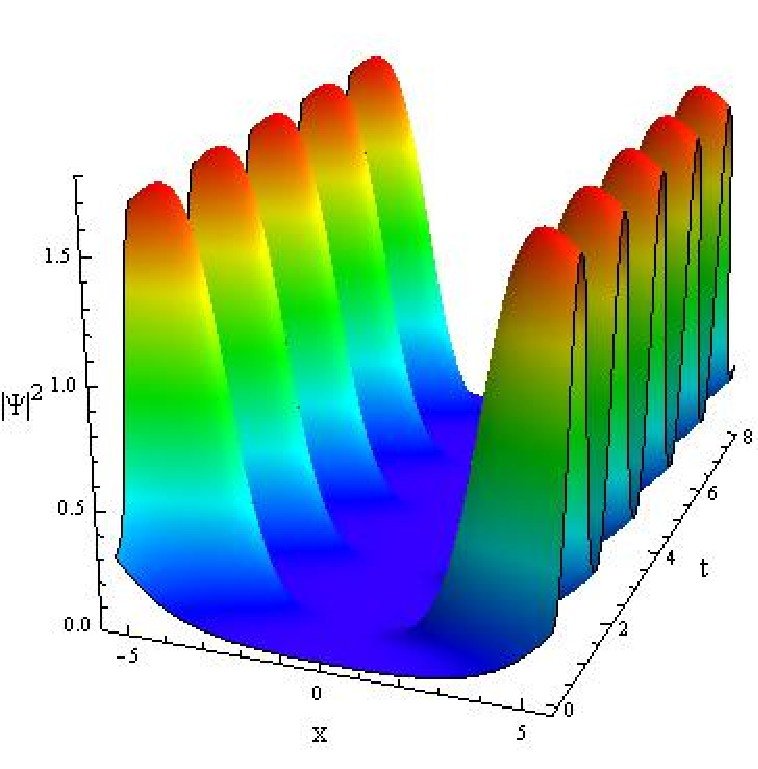}
\caption{Left: Profiles of $\left\vert \Phi \left( \zeta \right)
\right\vert ^{2}$ from Eq.(\ref{eq25a}). Right:$\left\vert \Psi
\left( x,t\right) \right\vert ^{2}$ for the wide breathing bright
soliton with $\lambda=0.001, b=10, a=0.01$, } \label{fig:1}
\end{figure}

We choose $\mu =2a^{2}-1\,,$ $\,G_{3}=2\left( a^{2}-2\right) \,,$ $%
\,G_{5}=3\,$ and $\epsilon =c=a^{2}$. The solution $\Phi \left(
\zeta \right) $, for the chosen coefficients of $U(\Phi )$, falls
into the first category of solutions shown in (\ref{eq20}) with
$g_{2}=\frac{4}{3}\left(
1+a^{2}\right) ^{2}$ and $g_{3}=-\frac{8}{27}\left( 1+a^{2}\right) ^{3}\,$%
and $\wp \left( \zeta ,g_{2},g_{3}\right) =\frac{1}{3}\left(
1+a^{2}\right) +\left( 1+a^{2}\right) \mathrm{csch}^{2}\left(
\sqrt{1+a^{2}}\zeta \right) \, $, such that we have\textrm{\ }
\begin{equation}
\Phi \left( \zeta \right) =\frac{a\,\tanh \left(
\sqrt{1+a^{2}}\,\zeta
\right) }{\sqrt{\mathrm{sech}^{2}\left( \sqrt{1+a^{2}}\zeta \right) +a^{2}\,}%
}\,.  \label{eq25a}
\end{equation}%
The profiles of $\left\vert \Phi \left( \zeta \right) \right\vert
^{2}$ for three different values of $\mu $ are shown in Fig.
(\ref{fig:1}), from which one can see the increasing of the width of
the soliton as $\mu $ becomes close to the critical value $\mu =-1$.

By using Eq. (\ref{foxt}) we find the wavefuncction
\begin{eqnarray}
\Psi \left( x,t\right) &=&\left. G_{3}^{1/6}\,\sqrt{\,\gamma }\exp \left( -%
\frac{\gamma ^{2}\,x^{2}}{6\,b^{2}}\right) \,\exp \left[ -i\,\eta
\left(
x,t\right) \right] \times \right.  \nonumber \\
&&\left. \times \frac{a\,\sinh \left( \sqrt{1+a^{2}}\,\frac{\sqrt{3\,\pi }\,b%
}{2\,G_{3}^{1/3}}\,\mathrm{Erfi}\left( \frac{\gamma
\,x}{\sqrt{3}\,b}\right) \right) }{\sqrt{1+a^{2}\,\cosh ^{2}\left(
\sqrt{1+a^{2}}\,\frac{\sqrt{3\,\pi
}\,b}{2\,G_{3}^{1/3}}\,\mathrm{Erfi}\left( \frac{\gamma \,x}{\sqrt{3}\,b}%
\right) \right) }}\right. ,  \label{25b}
\end{eqnarray}%
which is a wide breathing dark soliton for $a^{2}<1/2\,$.

\subsubsection{\textit{Example 2: }}

Bright soliton can also be modulate into a desired width and
amplitude in a controllable manner by changing the scattering length
and the trapping potential.

As in example 1, the soliton width can be modulated by a convenient
choice of the parameters. Here we take values of $\mu $ less than
the critical value $\mu =-1$ of the previous example and $U(\Phi )$
no longer positive,
but with two symmetrically disposed global minima and one local minima at $%
\Phi =0$. We look for solutions such that $\Phi (\zeta \rightarrow
\pm \infty )\rightarrow 0$. For that we take $\epsilon =c+k=0$,
where $k$ is a
constant of integration. From relations 18.12.1 to 18.12.3 of \ \cite%
{abramowitz} one can see that $\wp \left( \zeta ,4\mu ^{2}/3,8\mu
^{3}/27\right) =(\left\vert \mu \right\vert /3)(1+3~\mathrm{csch}^{2}\sqrt{%
\left\vert \mu \right\vert }\zeta )$ and by choosing $2\beta
/G_{3}=-\lambda
^{2}<0$ (that is, $G_{5}<3G_{3}^{2}/16\left\vert \mu \right\vert $) and $%
G_{3}<0$, such that $\sqrt{-\left\vert \mu \right\vert /G_{3}}$ is
real, we
find that%
\begin{equation}
\Phi \left( \zeta \right) =\frac{2\sqrt{-\left\vert \mu \right\vert /G_{3}}\,%
}{\sqrt{\lambda ^{2}\mathrm{\cosh }\left( 2\sqrt{\left\vert \mu \right\vert }%
\zeta \right) +1\,}}\,.  \label{25c}
\end{equation}%
\qquad\

\begin{figure}[h]
\centering
\includegraphics[width=7.0cm]{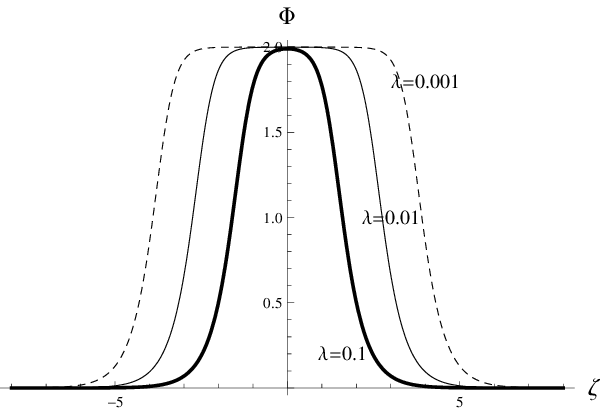}
\includegraphics[width=7.0cm]{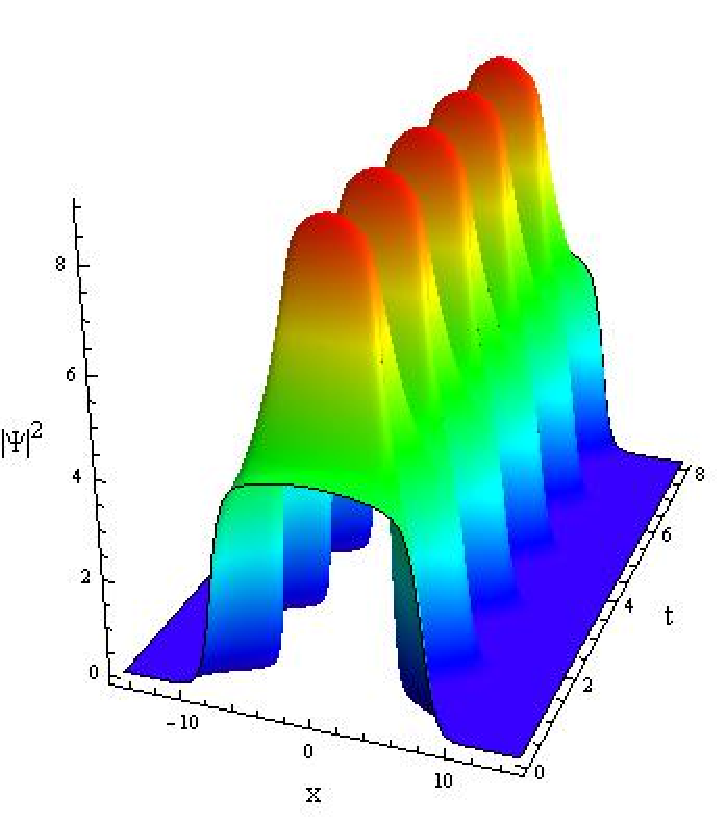}
\caption{Left: Profiles of $\left\vert \Phi \left( \zeta \right)
\right\vert ^{2}$ from Eq.(\ref{25c}). Right: $\left\vert \Psi
\left( x,t\right) \right\vert ^{2}$ for the wide breathing bright
soliton with $\lambda=0.001, b=10$}\label{fig:2}
\end{figure}

We show profiles of $\Phi \left( \zeta \right) $ for three different
values
of $~\lambda ^{2}<1$ and $\left\vert \mu \right\vert =-G_{3}=4~$in Fig. (%
\ref{fig:2}) . In the same figure we show also the amplitude
$\left\vert \Psi \left( x,t\right) \right\vert ^{2}$ for the wide
breathing bright soliton.

Particularly, a very thin bright soliton can be obtained when
$G_{5}=0$. We set $\mu =-1$ (the critical of $\mu $),
$\,G_{3}=-1\,$\ and $\epsilon =0\,$,
such that $g_{2}=\frac{4}{3}\,,$ $g_{3}=-\frac{8}{27},~\Delta =0\,\ $and $%
\Phi \left( \zeta \right) $ belongs to the second category of
solutions listed in the previous section, namely
\begin{equation}
\Phi \left( \zeta \right) =\left( \frac{1}{2}+\frac{1}{2}\left( \wp
\left( \zeta ,g_{2},g_{3}\right) -\frac{1}{3}\right) ^{-1}\right)
^{-1/2}\,, \label{23a}
\end{equation}%
One can verify that $\wp \left( \zeta ,g_{2},g_{3}\right) =\frac{1}{3}+%
\mathrm{csch}^{2}\zeta $\ and, consequently $\Phi \left( \zeta \right) =%
\sqrt{2}\,\mathrm{sech}\zeta $.

\noindent \qquad By using the Eq. (\ref{foxt}) we arrive at the
breathing bright soliton solution
\begin{equation}
\Psi \left( x,t\right) =\mathrm{e}^{i\pi /6}\sqrt{2\gamma }\exp \left( -%
\frac{\gamma ^{2}\,x^{2}}{6\,b^{2}}\right) \,\exp \left[ -i\,\eta
\left(
x,t\right) \right] \mathrm{sech}\left( -\frac{\sqrt{3\,\pi }\,b}{2\,}\,%
\mathrm{Erfi}\left( \frac{\gamma \,x}{\sqrt{3}\,b}\right) \right)
\,. \label{23b}
\end{equation}%
where $a(t)=\int \left( \gamma ^{2}\left( E-\frac{1}{3b^{2}}\right)
\right) dt.$

\subsubsection{\textit{Example 3: }}

This is an interesting example because we can find three different
analytic solutions with the same set of parameters $\mu =5\,,$
$\,G_{3}=10\,\ $and $\ G_{5}=-3\,$. Among those solutions we find
periodic solutions which were not presented in \cite{ABC}.

In the case $\epsilon =0$ and using (\ref{eq21}) we find two, very
similiar to each other, periodic solutions:
\begin{equation}
\Phi _{1}\left( \zeta \right) =\frac{\sqrt{10}}{\sqrt{\left( 5-\sqrt{5}%
\right) +2\sqrt{5}\sin ^{2}\left( \sqrt{5}\zeta \right) }}\,~~\mathrm{and~~}%
\Phi _{2}\left( \zeta \right) =\frac{\sqrt{10}}{\sqrt{\left( 5+\sqrt{5}%
\right) -2\sqrt{5}\sin ^{2}\left( \sqrt{5}\zeta \right) }}.
\label{24a}
\end{equation}%
However, if $\epsilon =\frac{5\left( 4\sqrt{10}-5\right) }{27}$ we
find the kink-like configuration

\begin{equation}
\Phi \left( \zeta \right) =\sqrt{\frac{4\sqrt{10}-5}{3}}\frac{\sinh
\left( \sqrt{\frac{5\left( \sqrt{10}-2\right) }{3}}\zeta \right)
}{\sqrt{3\left(
\sqrt{10}-2\right) +\left( \sqrt{10}+1\right) \sinh ^{2}\left( \sqrt{\frac{%
5\left( \sqrt{10}-2\right) }{3}}\zeta \right) }},  \label{24b}
\end{equation}%
which provides a breathing dark soliton solution for $\Psi \left(
x,t\right) $.

\section{Generalizations}

In this section we discuss some generalizations of the nonautonomous
nonlinear Schr\"{o}dinger equation. Particularly, we focus on two
generalizations.

\subsection{Generalization of the trapping potential}

As it was shown in equations (\ref{eq18b}), part of the potential is
determined by a convenient choice of $F\left( \xi \right) $, such that $%
F^{\prime }\left( \xi \right) $ does not have zeroes. Moreover, one
can see that $V\left( \xi \right) $ has, for $\mu =E=0$, a
strucuture similar to the one-dimensional supersymmetric quantum
mechanics potential, with $F^{\prime \prime }/2F^{\prime }$ playing
the role of supersymmetric superpotential. Then it would be
interesting to analyse the influence that several trapping
potentials in susy quantum mechanics could have while keeping the
same nonlinearities on the nonautonomous nonlinear Schr\"{o}dinger
equation.

We have analysed some of those possibilities and have found that the
trigonometric Scarf and trigonometric Rosen-Morse potentials which
have in their spectrum only bound-states (here we are using the same
nomenclature for susy quantum mechanics potentials as that of
\cite{khare}) are good candidates for the entrapment of breathing
bright solitons in a spatially periodic trapping potential.

For the sake of brevity we just present the first case here, that is we take%
\begin{equation}
F^{\prime }\left( \xi \right) =G_{3}^{-1/3}\,\sec ^{2A/\alpha
}\left( \alpha \xi \right) ,  \label{26}
\end{equation}%
where $A,\alpha >0$ and $-\pi /2\leq \alpha \xi \leq \pi /2$. Then
$V\left( \xi \right) =A\left( A-\alpha \right) \sec ^{2}\left(
\alpha \xi \right) -A^{2}-\mu \,G_{3}^{-2/3}\,\sec ^{4A/\alpha
}\left( \alpha \xi \right) $ and $g_{5}\left( x,t\right)
=G_{5}\,G_{3}^{-4/3}\,\sec ^{8A/\alpha }\left( \alpha \xi \right) $.

It is always convenient to look for $\gamma \left( t\right) $ and
$\delta \left( t\right) $ such that $\Psi \left( x,t\right) $ is
well defined. We choose
\begin{equation}
\gamma \left( t\right) =1+\left[ 1+\gamma _{1}\sin \left( t\right)
+\gamma _{2}\sin \left( \sqrt{2}t\right) \right] ^{2},\ \delta
\left( t\right) =0. \label{27}
\end{equation}%
We refer to \cite{ps2} for more details on this function $\gamma
\left(
t\right) $. The time-dependent frequency and driven force are given by (\ref%
{eq12}).

\begin{figure}[h]
\centering
\includegraphics[width=7.0cm]{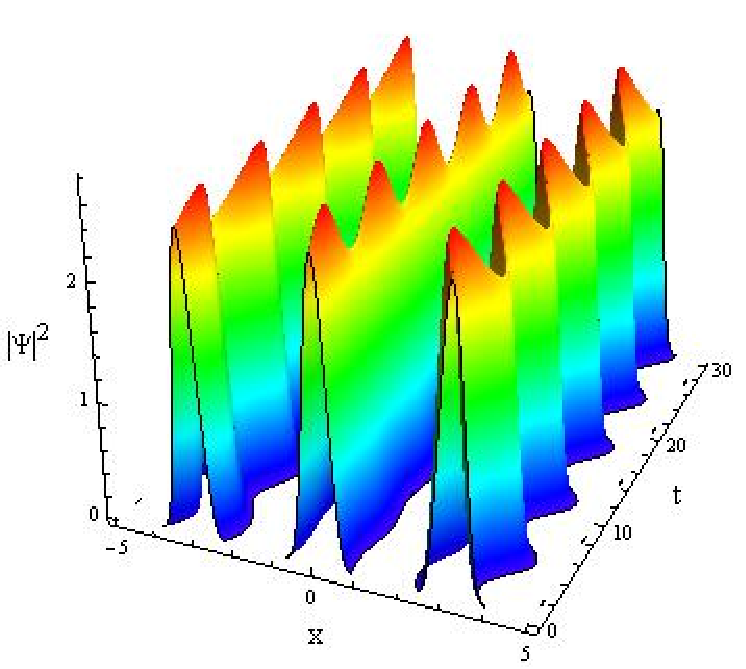}
\caption{$\left\vert \Psi \left( x,t\right) \right\vert ^{2}$ for
the bright soliton from Eq.(\ref{28}) with $\gamma_{1}=0,
\gamma_{2}=0.1$. The periodic structure is due to the periodicity of
the circular trapping potential.}\label{fig:3}
\end{figure}

In order to get explicitly brightlike solution, let us consider $\mu =0$, $%
G_{3}=2$, $G_{5}=-3$ and $\epsilon =0$. The solution is given by%
\begin{equation}
\Psi \left( x,t\right) =\frac{2^{1/6}\gamma ^{1/2}\exp \left[
-i\,\eta
\left( x,t\right) \right] \cos ^{A/\alpha }\left( \alpha \xi \right) }{\sqrt{%
1+F\left( \xi \right) ^{2}}},  \label{28}
\end{equation}%
whose modulus squared is shown in figure (\ref{fig:3}) for $A=\alpha
=1$.

A breathing dark solution is get by setting $\mu =3$, $G_{3}=6$,
$G_{5}=-3$
and $\epsilon =0$. It is given by%
\begin{equation}
\Psi \left( x,t\right) =\frac{6^{1/6}\gamma ^{1/2}\exp \left[
-i\,\eta \left( x,t\right) \right] F\left( \xi \right) \cos
^{A/\alpha }\left( \alpha \xi \right) }{\sqrt{1+F\left( \xi \right)
^{2}}}.  \label{29}
\end{equation}

\subsection{\noindent Generalization of the nonlinearities}

We have shown in equations (\ref{eq18b}) how the nonlinearities functions $%
g_{3}\left( x,t\right) $ and $g_{5}\left( x,t\right) $ (see Eqs. (\ref{eq1}%
)-(\ref{eq4})) must be intrinsically related to each other and given
in terms of the functions $\gamma \left( t\right) $ and $F^{\prime
}\left( \xi \right) $, such that the mapping of Eq. (\ref{eq1}) onto
(\ref{eq18}) could be realized. From a close inspection on those
relations we have found that the approach used here can be used to
map a nonautonomous Schr\"{o}dinger equation with non-polynomial
nonlinearity onto a static sine-Gordon like equation. In the first
paper in Ref. \cite{ps2} it was proposed a polynomial nonlinearity
up to Nth order in $\Psi (x,t)$, with fuctional coefficients related
to each order such that the nonautonomous NLSE could be mapped onto
a stationary NLSE with constants coefficients. Here we present
inverse transformations to obtain a nonautonomous NLSE with
non-polynomial nonlinearity from a stationary NLSE with
non-polynomial nonlinearity.
Particularly, we consider the sine-Gordon equation which is satisfied by $%
\Phi (\zeta )$, namely%
\begin{equation}
\frac{d^{2}\Phi \left( \zeta \right) }{d\zeta ^{2}}=\frac{1}{b^{2}}\sin %
\left[ b\,\Phi \left( \zeta \right) \right] =\sum_{n=0}^{\infty }\frac{%
\left( -1\right) ^{n}}{\left( 2n+1\right) !}b^{2n-1}\Phi ^{2n+1},
\label{30}
\end{equation}%
where $b$ is positive constant. By using the transformation $\zeta
=F\left( \xi \right) $ and the redefinition of $\Phi \left( F\left(
\xi \right) \right) =\sqrt{F^{\prime }\left( \xi \right) }\,\phi
\left( \xi \right) $ we
find that%
\begin{equation}
\frac{d^{2}\phi }{d\xi ^{2}}=\left( \left( \frac{F^{\prime \prime }}{%
2F^{\prime }}\right) ^{2}-\left( \frac{F^{\prime \prime }}{2F^{\prime }}%
\right) ^{\prime }\right) \phi +\sum_{n=0}^{\infty }\frac{\left(
-1\right) ^{n}}{\left( 2n+1\right) !}(F^{\prime
}\,)^{n+2}b^{2n-1}\,\phi ^{2n+1}. \label{31}
\end{equation}%
Now, by considering that $\phi \left( \xi \right) =\psi \left( \xi
,\tau \right) {e}^{iE\,\tau }$, we find that $\Psi \left( \xi ,\tau
\right) $
satisfies the nonautonomous NLSE%
\begin{equation}
i\frac{\partial \psi }{\partial \tau }=-\frac{\partial ^{2}\psi
}{\partial \xi ^{2}}+V\left( \xi \right) \psi +\sum_{n=0}^{\infty
}\frac{\left( -1\right) ^{n}}{\left( 2n+1\right) !}(F^{\prime
}\,)^{n+2}\,b^{2n-1}\,|\psi |^{2n}\psi ,  \label{32}
\end{equation}%
where $V\left( \xi \right) =\left( \frac{F^{\prime \prime }}{2F^{\prime }}%
\right) ^{2}-\frac{d}{d\xi }\left( \frac{F^{\prime \prime }}{2F^{\prime }}%
\right) +E$.

Now, the inverse of transformation in equations
(\ref{eq5})-(\ref{eq6}), namely
\begin{equation}
\xi =\gamma \left( t\right) x+\delta \left( t\right)
\,~\mathrm{and~}\tau -\tau _{0}=\int_{0}^{t}dt^{\prime }\gamma
^{2}\left( t^{\prime }\right) \,, \label{33}
\end{equation}

lead us to
\begin{eqnarray}
\frac{i}{\gamma ^{2}}\frac{\partial \,\psi }{\partial \,\tau }
&=&\left. \frac{i}{\gamma ^{3}}\left( \gamma _{t}\,x+\delta
_{t}\right) \frac{\partial
\,\psi }{\partial \,x}-\frac{1}{\gamma ^{2}}\frac{\partial ^{2}\,\psi }{%
\partial \,x^{2}}+V\left[ \xi \left( x,t\right) \right] \,\psi +\right.
\nonumber \\
&&+\sum_{n=0}^{\infty }\frac{\left( -1\right) ^{n}}{\left( 2n+1\right) !}%
b^{2n-1}\,(F^{\prime }\,)^{n+2}\,|\psi |^{2n}\psi ,  \label{34}
\end{eqnarray}%
where $\psi =\psi (\xi \left( x,t\right) ,\tau \left( t\right) )$.
The first-derivative in $x$ can be eliminated by the following
redefinition of the wavefunction$~\psi (x,t)=e^{i\,\alpha \left(
x,t\right) }\Psi \left( x,t\right) /\sqrt{\gamma \left( t\right) }$,
where $\alpha \left( x,t\right) =\frac{\gamma _{t}}{4\gamma
}x^{2}+\frac{\delta _{t}}{2\gamma }x-a\left( t\right) $. Then we
finally obtain the nonautonomous NLSE with a generalized
nonlinearity

\begin{equation}
i\frac{\partial \Psi }{\partial t}=-\frac{\partial ^{2}\Psi }{\partial x^{2}}%
+\upsilon \left( x,t\right) \,\Psi +\sum_{n=0}^{\infty }\frac{\left(
-1\right) ^{n}}{\left( 2n+1\right) !}b^{2n-1}\,\left( \frac{F^{\prime }}{%
\gamma }\right) ^{n+2}\,\,|\Psi |^{2n}\Psi \,=\frac{\delta \mathbf{H}}{%
\delta \Psi ^{\ast }(x,t)}  \label{35}
\end{equation}%
where
\begin{eqnarray}
\upsilon \left( x,t\right) &=&\left. \frac{\gamma _{tt}\gamma
-2\gamma _{t}^{2}}{4\gamma ^{2}}\,x^{2}+\frac{\delta _{tt}\gamma
-2\gamma _{t}\delta
_{t}}{2\gamma ^{2}}\,x\right.  \nonumber \\
&&\left. -\frac{\delta _{t}^{2}}{4\gamma ^{2}}-\frac{da}{dt}+\gamma ^{2}\,V%
\left[ \xi \left( x,t\right) \right] \,\right.
\end{eqnarray}%
and

\begin{eqnarray}
\mathbf{H} &=&\int dz~\left\{ \Psi ^{\ast }(z,t)\left[ -\frac{\partial ^{2}}{%
\partial z^{2}}+\upsilon \left( z,t\right) \,\right] \Psi (z,t)\right.
\nonumber \\
&&\left. +\frac{2}{b^{3}}\gamma ^{3}F^{\prime }\left[ \xi \left( z,t\right) %
\right] \,\cos \left( b\sqrt{\frac{F^{\prime }\left[ \xi \left( z,t\right) %
\right] }{\gamma }}|\Psi |\right) \right\} .
\end{eqnarray}

In this specific generalization the soliton amplitude is periodic
and given by
\begin{equation}
\left\vert \Psi \left( x,t\right) \right\vert ^{2}=\frac{4\gamma
\left( t\right) ~\left[ \arccos \left( \nu \,\mathit{sn}\left( F(\xi
)/\sqrt{b}|\nu \right) \right) \right] ^{2}}{b^{2}F^{\prime }(\xi
)},
\end{equation}%
where $\mathit{sn}\left( z|\nu \right) $ is the snoidal Jacobi
elliptic
function with elliptic parameter $0\leq \nu \leq 1$. For $\nu =1$, $\mathit{%
sn}\left( z|\nu \right) =\tanh (z)$ and we have a breathing dark
soliton.

\section{Conclusions}

In this work we have shown that an approach used to tackle the Schr\"{o}%
dinger equation with explicit time-dependent potential parameters
can be extend to the case of non-autonomous NLSE. As a consequence
we were able to reproduce results originally obtained through the
use of an Ansatz together with similarity transformation. Besides,
this procedure lead us to novel solutions of that problem and,
particularly some configurations of wide soliton solutions.

Furthermore, we extended the approach in order to deal with systems
with generalized nonlinearities and trapping potentials which are a
mixing of the (driven) time-dependent harmonic oscillator and
circular functions. Extensions of this work for the cases with two
or more fields are under development.

\section{Acknowledgments}

LEAM thanks to Brazilian funding agency CAPES for the scholarship
under PEC-PG program. This work is also partially supported by CNPq
(procs. 482043/2011-3, 304352/2009-8 and 304252/2011-5 ).

\end{document}